\documentclass[twocolumn,showpacs,preprintnumbers,amsmath,amssymb,prb]{revtex4}

\usepackage{graphicx}
\usepackage{bm}

\begin{document}

%\preprint{APS/123-QED}

\title{Temperature and Field Dependence of the Energy Gap of MgB$_2$/Pb planar junction}

\author{Mohamed H. Badr$^{1,2}$, Mario Freamat$^1$, Yuri Sushko$^1$, and K.-W. Ng$^1$} 
\affiliation{
$^{1}$Department of Physics and Astronomy, University of Kentucky, Lexington, KY 40506-0055, U.S.A.\\
$^{2}$Department of Physics, Faculty of Science, Menoufiya University, Menoufiya, Egypt.}

\date{\today}

\begin{abstract}
We have constructed MgB$_2$/Pb planar junctions for both temperature and field 
dependence studies. Our results show that the small gap is a true bulk 
property of MgB$_2$ superconductor, not due to surface effects. The temperature
dependence of the energy gap manifests a nearly BCS-like behavior. Analysis 
of the effect of magnetic field on junctions suggests that the energy gap of
MgB$_2$ depends non-linearly on the magnetic field. Moreover, MgB$_2$ has an 
upper critical field of 15 T, in agreement with some reported $H_{c2}$ from 
transport measurements. 
\end{abstract}

\maketitle

\section{\label{sec:level1}Introduction\protect\\}

Recently Nagamatsu et al.\cite{Nagam} has discovered superconductivity in the
commonly available compound MgB$_2$ with a $T_c$ of 40 K. Similar to the 
cuprates, MgB$_2$ has a layer structure and hence many of its superconducting
properties may show anisotropic effect. For instance, the anisotropy ratio 
$\gamma=\xi_{ab}/\xi_c$ has a reported value that varies from 1.1 to 9.0 
\cite{Hands,Simon}. There is also evidence that the energy gap 
value is very different along these two directions showing anisotropic s-wave 
or two-gap behavior\cite{Plece}$^-$\cite{Laube}. On the other hand, as clearly
demonstrated by isotope effect \cite{Budko,Hinks} and neutron 
scattering\cite{Osbor,Yildi}, MgB$_2$ is different from the cuprates and its 
Cooper pairs are phonon mediated.

Although the pairing mechanism in MgB$_2$ is thought to be phonon mediated,
there are still many experimental results that lack appropriate explanation.
Many of these unanswered problems may lead to unexpected and interesting
physics of superconductivity. For example, there is no consensus about the
magnitude of the energy gap until now. Many techniques have been used to
measure the gap such as Raman spectroscopy\cite{Chen02,Quilt}, far-infrared
transmission\cite{Kaind}$^-$\cite{Gorsh}, specific
heat\cite{Bouqu}$^-$\cite{Bauer}, high-resolution photoemission\cite{Takah}
and tunneling\cite{Plece}$^-$\cite{Laube,Rubio}$^-$\cite{Gonne}.
Most tunneling data on MgB$_2$, as in the case of many other newly discovered
superconductors, are obtained from mechanical junctions like scanning tunneling
microscope\cite{Chen01,Giubi01,Giubi02,Rubio}$^-$\cite{Sharo}, point contact
\cite{Szabo,Laube,Schmi}$^-$\cite{Gonne} and tunneling junctions\cite{Plece}.

It is critical to determine whether the small gap value reported by many
groups\cite{Rubio} is a real bulk property or a result of surface degradation. 
One direct method is to measure the temperature dependence of the energy gap. 
Since the structure of a mechanical junction will change as the temperature is
varied or an external field is applied, it is not stable enough to study 
temperature dependence of the energy gap. The situation will be worse if the 
sample is not homogeneous and the gap value varies with the probe position.   
The more reliable measurement for temperature dependence of the energy gap is 
from sandwich type planar junctions where any variation in the tunneling 
spectra will be a pure result of the sample under study not due to any 
structural changes in the junction. To our best knowledge, in this paper we 
report the first energy gap temperature and magnetic field dependence of 
MgB$_2$ by planar junctions.

\section{Experimental Details}
MgB$_2$ sample is prepared by reacting Mg turnings (99.98\%) and boron powder
(99.99\%, -325 mesh) with the stoichiometric composition 1:2 respectively. 
Magnesium and boron are mechanically pressed and sealed in a tantulum tube 
(99.9\%, 2.4 mm inner diameter). The tantalum tube is then sealed inside a 
quartz tube and placed inside a box furnace at 950 $^{\circ}$C. It is then
quenched to room temperature after two hours. The polycrystalline MgB$_2$ is 
then characterized using X-ray diffraction, resistivity and dc SQUID 
magnetometer (Quantum Design MPMS) measurements.

Junctions are constructed by attaching two leads to MgB$_2$ sample and molding
it inside epoxy resin. It is then ground to expose the sample and mechanically 
polished to a smoothness of 0.3 micron. Pb, a superconductor with  
$T_c\approx$ 7.2 K and $H_c(0)\approx$ 0.08 T, is evaporated on the top as
a counter electrode. We used Pb to sharpen the peak features and also as a 
control to monitor the tunneling conditions. In this paper we will limit our 
analysis only to the data when Pb is normal, which is simpler to understand. 
We have also attempted to grow artificial barrier by sandwiching a thin 
oxidized aluminum layer between the sample and Pb electrode. This will in 
general lead to very large junction resistance, even with the minimal 
thickness of aluminum layer. So far, the best junctions are still from those 
with natural barrier. The junctions show stability against any temperature
changes in the full range from 4.2 K to room temperature, but can survive only 
up to a magnetic field (perpendicular to the barrier) of approximately 3.2 T.

\section{Results and discussions}
\begin{figure}
\includegraphics{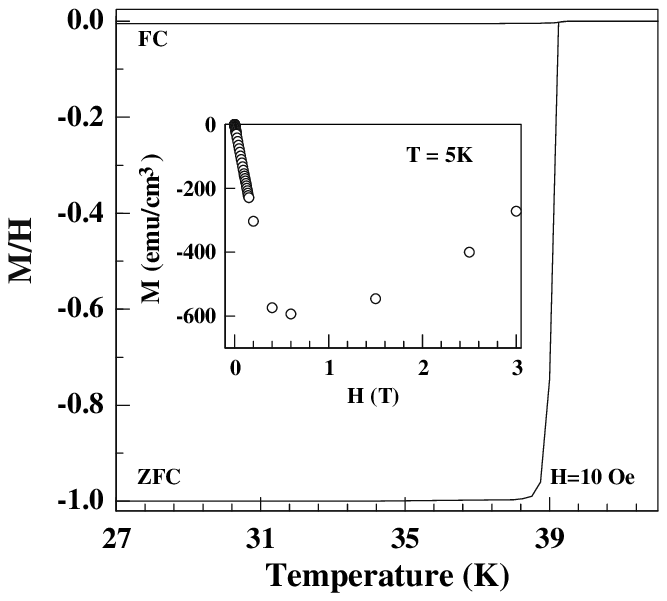}
\caption{\label{fig:F1} Magnetization divided by an applied filed of 10 Oe as
a function of temperature for both Zero Field Cooled (ZFC) and Field Cooled 
(FC) modes. The inset shows the magnetization curve $M(H)$ measured at 
$T=$ 5K.  For clarity, the x-axis (inset) is limited to small values of $H$.}
\end{figure}

X-ray diffraction pattern shows no trace of other phases in the sample. 
Both resistivity and SQUID measurements have determined the 
$T^{\text{onset}}_c$ to be 39.5 K (defined by 2\% criteria) with a sharp 
transition width of 0.7 K (10\%-90\% criteria). Fig.~\ref{fig:F1} (main panel)
shows the temperature dependence of susceptibility  for both zero-field cooled 
(ZFC) and field cooled (FC) modes. Taking in account the demagnetizing factor 
of the measured cylindrical  sample with $\gamma=$ 1 (ratio of length to 
diameter), the sample shows a perfect diamagnetic shielding $M/H=-$ 1. This 
result along with a Residual Resistance Ratio $RRR=R(300)/R(T_c)=$ 8 reflects 
both the sample's good quality and grains coupling. The small FC susceptibility
signal observed here is a common feature for such polycrystalline MgB$_2$ 
samples sintered around 950 $^{\circ}$C or higher\cite{Takan,Glowa}. This can 
be attributed to large trapping of flux by cracks and voids that reflects also 
the good coupling of grains\cite{Paran}. It is interesting that the lower
critical field $H_{c1}(5K)=$ 0.2 T as estimated from the magnetization curve 
(Fig.~\ref{fig:F1}, inset). This value is significantly larger than those 
reported by other research groups\cite{Takan,Joshi} and can be attributed to
the good quality of the sample.

\begin{figure}
\includegraphics{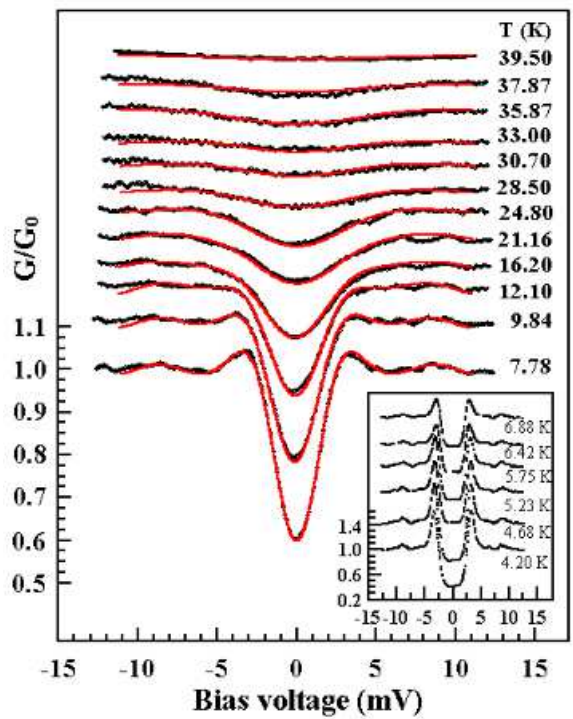}
\caption{\label{fig:F2} Temperature dependence of the experimental tunneling 
conductance spectra normalized by the conductance at 15 mV (dots). The spectra
at temperatures above $T_{c}$ of lead are represented in the main panel with 
the corresponding two-gap BTK fittings (red lines). For clarity the curves are 
vertically shifted.}
\end{figure}

The inset of Fig.~\ref{fig:F2} shows how the conductance curves evolve with 
temperatures below lead $T_c$. For this S/I/S junction we can roughly estimate 
the energy gap of MgB$_2$. It is clear that the spectra are sharpened 
significantly as the Pb gap $\Delta_{Pb}$ opens up. Since $\Delta_{Pb}(0)$ is 
about 1.2 meV and the peak position of the 4.2K curve is at 3.2 meV which can 
be considered as the half sum of gaps, we can estimate $\Delta_{MgB_2}$ to 
be about 2 meV. The corresponding value of $2\Delta/kT_c$ is only about 1.18, 
much smaller than the BCS value of weak coupling superconductors. This is 
consistent with other small gap results from tunneling measurements\cite{Rubio}.
As can be seen from Fig.~\ref{fig:F2} (main panel), the smaller peak around 9 
meV survives for temperatures up to 21.16 K. Similar features are commonly 
found in other tunneling data\cite{Giubi01}.

It is intriguing to find this very small energy gap in a high $T_c$ 
superconductor like MgB$_2$. All tunneling data published so far can be 
summarized into three main categories, according to the interpretation: 
one-gap\cite{Rubio}$^-$\cite{Schmi,Gonne}, 
two-gap\cite{Plece,Giubi01}$^-$\cite{Laube} and gap anisotropy\cite{Chen01}.
If there is only one single gap, then it is likely that the small value is
a result of surface degradation. However, our data do not support this 
explanation because the gap exists up to the bulk $T_c$. So far, there is no 
direct observation of two distinguished gaps in the same tunneling spectrum. 
Mostly a small feature at a higher energy is interpreted as the second 
gap\cite{Szabo}. This two-gap approach is in accordance with the 2D and 3D 
Fermi surfaces proposed by Liu et al.\cite{Liu}. On the other hand, Chen et 
al.\cite{Chen01} proposed an anisotropic s-wave pairing model with 
$\Delta_{xy}=$ 5 meV and $\Delta_z=$ 8 meV to best fit their tunneling curves.

We have used Blonder, Tinkham, and Klapwijk\cite{BTK} (BTK) model to analyze 
the curves when Pb is normal (Fig.~\ref{fig:F2}, main panel). In this model,
in addition to quasiparticle tunneling, the possibility of Andreev reflection 
and normal reflection by the barrier is also included. As indicated by our 
fitting, the barrier strength Z of our junction is not strong enough to prevent
Andreev reflection from happening. A depairing term $\Gamma$ is also included 
because of shortening in quasiparticle lifetime by different scattering processes.

Here we have assumed the two-gap model to cover also the small feature around 
9 meV (Fig.~\ref{fig:F2}, main panel). We consider this feature as the second 
energy gap $\Delta_2$ in MgB$_2$. Also, the two gaps contribute to tunneling
independently. Therefore, we assign a small percentage of tunneling $C_2$ for 
$\Delta_2$ and $C_1=1-C_2$ for the smaller gap $\Delta_1$. The parameters  
$\Gamma$, Z, $\Delta_1, \Delta_2, C_1$ and $C_2$ are used to best fit the 
curve at 7.78 K and their values are 0.95 meV,  1.33,  1.75 meV, 8.2 meV, 0.94 
and 0.06, respectively. All these parameters except the $\Delta$'s are kept 
constant for all higher temperature curves, i.e., the $\Delta$'s are the only 
adjusting parameters. From the quality of the fittings, it is justifiable to 
say that $\Gamma$, Z and $C$'s are independent of energy and temperature within 
the ranges of our measurement. Furthermore, the zero bias offset is purely a 
result of Andreev reflection at the barrier. 

\begin{figure}
\includegraphics{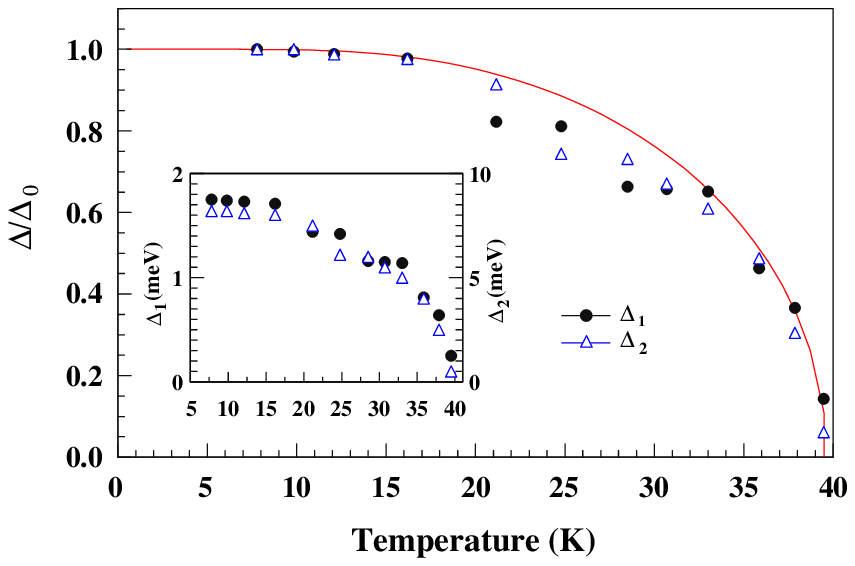}
\caption{\label{fig:F3} Temperature dependence of the two gaps $\Delta_1$ and
$\Delta_2$ normalized by their values at $T=$ 7.78 K. The absolute values are 
shown in the inset. The solid line represents the expected BCS 
$\Delta(T)/\Delta(0)$ with $T_c=$ 39.5 K.}
\end{figure}

The inset of Fig.~\ref{fig:F3} shows the temperature dependence of the two 
gaps. The main panel shows both gaps normalized to their values at $T=$ 7.78 K 
along with the BCS expected behavior (solid line). As can be seen, both 
$\Delta_1$ and $\Delta_2$ survive to $T_c$ of bulk MgB$_2$ with a small 
deviation from BCS as we try to apply the two-gap model. Since the tunneling 
features are mainly due to $\Delta_1$ ($C_1=$ 0.94) and they survive up to 
$T_c$ of MgB$_2$, it is justifiable to consider $\Delta_1$ as a true bulk 
property of this superconductor. This two-gap fitting gives 
$\Delta_2(0)/\Delta_1(0)\approx$ 4.5, close to both the theoretically 
predicted\cite{Liu} and experimentally suggested\cite{Bouqu} value. 
Nevertheless, there are still unexplained problems with this two-gap model. 
For example, why the large gap contributes that little to tunneling,
$C_2=$ 0.06, for such a polycrystalline sample?  Our analysis above still 
holds for a single gap superconductor by setting $C_2=$ 0, but then we have 
to explain why $\Delta_1$ is so small. 

\begin{figure}
\includegraphics{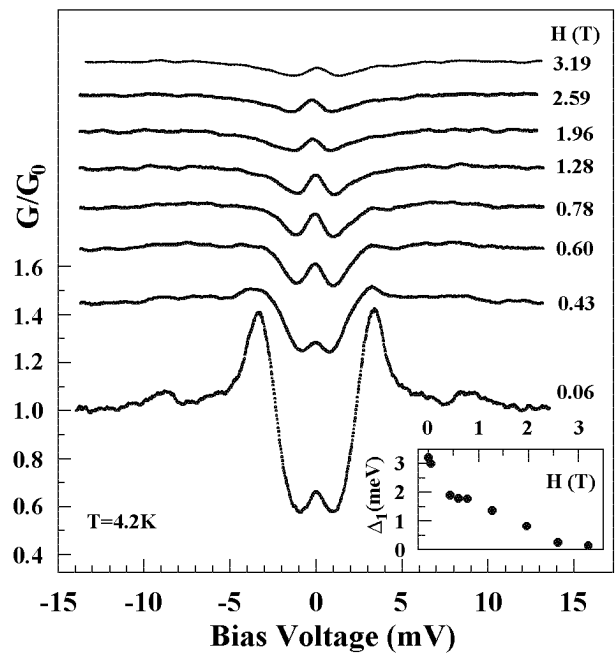}
\caption{\label{fig:F4} Magnetic field dependence of the experimental tunneling 
conductance spectra normalized by the conductance at 15 V. The inset shows 
the field dependence of $\Delta_1$ measured at 4.2 K.}
\end{figure}

To further characterize the junctions, we have also studied the field 
dependence of the tunneling spectra at 4.2 K (Fig.~\ref{fig:F4}, main panel). 
The junction in an external field normal to the barrier  is not as stable as
its performance against temperature changes. It experiences slight changes even
when a small field is applied. This can be seen from the development of the 
zero bias conductance peak, similar to that observed by another 
group\cite{Schmi}. This can be explained by enhancement of micro-shorts through 
the barrier as a result of the applied field. Furthermore, most junctions 
collapse and the tunneling spectra transit from quasiparticle to Josephson 
tunneling at fields of about 3.2 T. In this paper, we focus only on 
quasiparticle tunneling spectra. It can be seen from Fig.~\ref{fig:F4} 
(main panel) that the quality of the spectra has severely degraded when $H$ 
exceeded $H_c$ of lead. The curve at 0.43 T is more severely smeared and 
depressed as compared to the curve at $T=$ 7.78 K (Fig.~\ref{fig:F2}). This 
reflects the fact that this field is already greater than $H_c$ of Pb. Using 
the peak position of the small gap $\Delta_1$, we can roughly estimate its 
dependence on $H$ as shown on Fig.~\ref{fig:F4} (inset). It is worth noting that 
$\Delta_1(0T)-\Delta_1(0.43T)>\Delta_{Pb}$. This supports the above argument
on the condition of the curve at 0.43 T. Moreover, the energy gap depends
non-linearly on the the magnetic field. Further work should be done to 
investigate this dependence.

\begin{figure}
\includegraphics{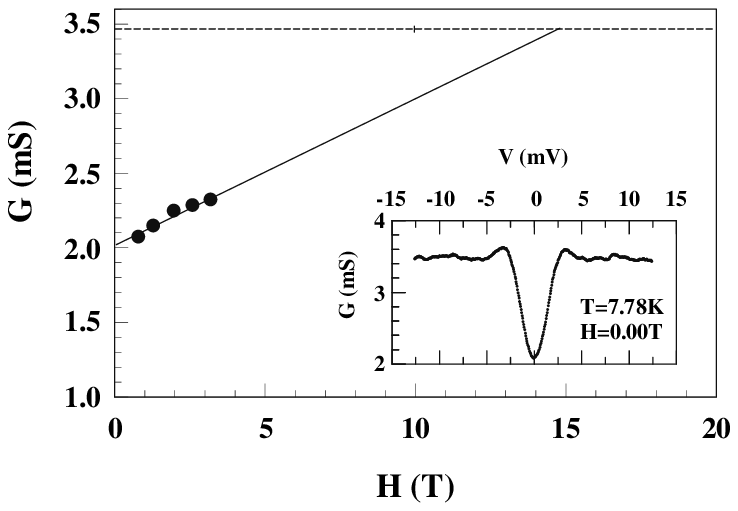}
\caption{\label{fig:F5} Magnetic field dependence of the minimum conductance.
The linear fit intersects the normal conductance line in a point corresponding
to H$_{c2}\approx$ 15 T. The intersection with the vertical axis matches the 
minimum conductance offset of the spectrum at 7.78 K and 0 T (inset).}
\end{figure}

Since MgB$_2$ is a type II superconductor, the effect of magnetic field is to 
produce vortices and hence the order parameter on the surface is not 
homogeneous anymore. In a simple model, the tunneling spectrum is an ensemble 
of all different gap values sampled within the junction area (0.05 mm$^2$). 
If we consider the vortex core as a normal region, and the number of vortices 
produced is proportional to the applied normal  field, then the zero bias 
conductance should be proportional to that field\cite{Colli}. To study this 
dependence, we have fitted our tunneling spectra within the gap by parabola to 
remove the zero conductance peak. The zero conductance can then be estimated 
from the parabola. In Fig.~\ref{fig:F5} (main panel) we have plotted the zero 
bias conductance versus the external applied field. It is clear that the zero 
bias offset increases linearly with the external field. By extrapolation to the 
normal conductance, we can estimate $H_{c2}$ of MgB$_2$ to be about 15 T.  
This value is in agreement with $H_{c2}$ of bulk MgB$_2$ from transport 
measurements (see, e.g., Ref. 29, 36) rather than the small reported value 
of about 6 T from tunneling analysis (see, e.g., Ref. 24). From 
Fig.~\ref{fig:F5} we can also estimate the S/I/N zero bias offset at zero field
to be around 2.1 mS. This agrees with the zero bias offset in the zero field 
conductance curve at 7.78 K (Fig.~\ref{fig:F5}, inset).  

\section{Conclusion}
We have prepared MgB$_2$/Pb planar junctions to study the temperature and field 
dependence of the energy gap of MgB$_2$. The temperature dependence data 
indicate that the small energy gap we have measured is indeed a bulk property 
of MgB$_2$. Moreover, our data do not contradict the two-gap scenario by 
considering our reported gap of about 2.0 meV as the small gap. Analysis of the
effect of magnetic field on the junctions shows that MgB$_2$ has an upper 
critical field of about 15 T which is consistent with most transport 
measurements of H$_{c2}$. Moreover, the energy gap shows a non-linear 
dependence on the magnetic field applied perpendicular to the barrier.

This work is supported by NSF Grant No. DMR997201. 

% The Reference section begins

\setlength{\textwidth}{7.1in}
\setlength{\textheight}{9.7in}
\small \rm

\end{document}